\documentclass[11pt,a4paper]{article}

\usepackage[final]{graphics}
\usepackage[dvips]{graphicx}
\input{epsf}
\usepackage[intlimits]{amsmath}
\usepackage{amsfonts}
\usepackage{booktabs}
\usepackage{epsfig}
\usepackage{url}
\usepackage{fancyvrb}

\def\line#1{\hbox to \textwidth{#1}}
\catcode`\@=11

\def\thebibliography#1{\section*{BIBLIOGRAPHY}\list{\arabic{enumi}.}
  {\settowidth\labelwidth{#1.}\leftmargin=1.67em
   \labelsep\leftmargin \advance\labelsep-\labelwidth
   \itemsep\z@ \parsep\z@
   \usecounter{enumi}}\def\makelabel##1{\rlap{##1}\hss}%
   \def\newblock{\hskip 0.11em plus 0.33em minus -0.07em}
   \sloppy \clubpenalty=4000 \widowpenalty=4000 \sfcode`\.=1000\relax}

\def\@cite#1#2{$[{{#1\if@tempswa , #2\fi}}]$}

\newcount\@tempcntc
\def\@citex[#1]#2{\if@filesw\immediate\write\@auxout{\string\citation{#2}}\fi
  \@tempcnta\z@\@tempcntb\m@ne\def\@citea{}\@cite{%
	\@ordonner{#2}%
	\@for\@citeb:=#2\do%
    {\@ifundefined{b@\@citeb}%
	{\@citeo\@tempcntb\m@ne\@citea%
        	\def\@citea{,\penalty\@m\ }{\bf ?}\@warning%
       		{Citation `\@citeb' on page \thepage \space undefined}}%
    	{\setbox\z@\hbox{\global\@tempcntc0\csname b@\@citeb\endcsname\relax}
     \ifnum\@tempcntc=\z@ \@citeo\@tempcntb\m@ne%
       \@citea\def\@citea{,\penalty\@m}%
       \hbox{\csname b@\@citeb\endcsname}%
     \else%
      \advance\@tempcntb\@ne%
      \ifnum\@tempcntb=\@tempcntc%
      \else\advance\@tempcntb\m@ne\@citeo%
      \@tempcnta\@tempcntc\@tempcntb\@tempcntc\fi\fi}}\@citeo}{#1}}%

\def\@citeo{\ifnum\@tempcnta>\@tempcntb\else\@citea
  \def\@citea{,\penalty\@m}%
  \ifnum\@tempcnta=\@tempcntb\the\@tempcnta\else
   {\advance\@tempcnta\@ne\ifnum\@tempcnta=\@tempcntb \else
\def\@citea{-}\fi
    \advance\@tempcnta\m@ne\the\@tempcnta\@citea\the\@tempcntb}\fi\fi}

\def\@toto{}
\newif\if@ordre 
\newcount\c@current
\newcount\c@last

\def\@ordonner#1{\global\c@last\m@ne%
		\global\@ordretrue%
		\@for\@toto:=#1\do%
			{\@ifundefined{b@\@toto}%
			{}%
			{\c@current\csname b@\@toto\endcsname\relax%
			\ifnum\the\c@current<\the\c@last\relax%
				{\global\@ordrefalse}\fi%
			\global\c@last\the\c@current%
			}%
			}%
		\if@ordre{}\else{\typeout{}%
			\typeout{Warning: the references are not %
			 in increasing order\on@line:}%
			\@for\@toto:=#1\do%
			{\@ifundefined{b@\@toto}%
			{}%
			\typeout{\@toto:\space \@nameuse{b@\@toto}}%
			}\typeout{}}\fi%
		}%

\catcode`\@=12

\DeclareFontFamily{OT1}{chan}{}
\DeclareFontShape{OT1}{chan}{m}{n}{<-> s * [1.08] pzcmi}{}
\newcommand{\xloops}{XLOOPS}
\newcommand{\ginac}{{GiNaC}}
\newcommand{\maple}{{\sffamily Maple}}
\newcommand{\mathematica}{{\sf Mathematica}}
\newcommand{\mupad}{{\sf MuPAD}}
\newcommand{\reduce}{{\sf Reduce}}
\newcommand{\cln}{{\sf CLN}}

\DeclareMathOperator{\Tr}{Tr}

\newcommand*{\CC}{C${}^{{}_{++}}$}

\newcommand{\lslash}{l\hspace{-1.0ex}/}  
 
\newcommand{\qslash}{q\hspace{-1.0ex}/} 

\newcommand{\intddl}{\int \!\! d^Dl}

\newcommand{\slL}{\raise.15ex\hbox{$/$}\kern-.53em\hbox{$L$}}
\newcommand{\slP}{\raise.15ex\hbox{$/$}\kern-.53em\hbox{$P$}}
\newcommand{\slR}{\raise.15ex\hbox{$/$}\kern-.53em\hbox{$R$}}
\newcommand{\slQ}{\raise.15ex\hbox{$/$}\kern-.53em\hbox{$Q$}}
\newcommand{\slK}{\raise.15ex\hbox{$/$}\kern-.53em\hbox{$K$}}

\newenvironment*{codeblock}%
  {\Verbatim[fontsize=\small,%
             numbers=left,%
             baselinestretch=0.9,%
             xleftmargin=2em,%
             numbersep=1em,
             commandchars=\\\{\}]}
  {\endVerbatim}

\def\build#1\over#2{\mathrel{\mathop{\kern 0pt#2}\limits_{#1}}}

\font\tenimbf=cmmib10 at 12pt
\font\sevenimbf=cmmib10 at 7pt
\font\fiveimbf=cmmib10 at 5pt
\newfam\imbf
\textfont\imbf=\tenimbf
\scriptfont\imbf=\sevenimbf
\scriptscriptfont\imbf=\fiveimbf

\begin{document}
\title{
\vskip-3cm
{\baselineskip14pt \normalsize
\hfill MZ-TH/01-02 }
\vskip4cm                
One-loop integrals with \xloops-\ginac\footnote{Supported by the Deutsche Forschungsgemeinschaft and the "Graduiertenkolleg Eichtheorien" at the University of Mainz.}\\[3ex]
\author{C.~Bauer, H.S.~Do\\[2ex]
\normalsize{Institut f\"ur Physik, Johannes-Gutenberg-Universit\"at,}\\
\normalsize{Staudinger Weg 7, D-55099 Mainz, Germany} \\[10ex]
} }
\date{}

\maketitle 

\begin{abstract}
\medskip
\noindent
We present a new algorithm for the reduction of one-loop tensor Feynman
integrals within the framework of the \xloops\ project, covering both
mathematical and programming aspects. The new algorithm supplies a clean way
to reduce the one-loop one-, two- and three-point Feynman integrals with
arbitrary tensor rank and powers of the propagators to a basis of simple
integrals.
\noindent
We also present a new method of coding \xloops\ in \CC\ using the \ginac\
library.
\end{abstract}

\thispagestyle{empty} 

\newpage

\section{Introduction}
The purpose of this paper is twofold. First, we introduce a new method to
code \xloops\ in \CC\ with the help of the \ginac\ library~\cite{ginac}. This
also serves as an introduction to \ginac\ for symbolic computations.

Second, we introduce a new procedure for tensor reduction which allows us to
manipulate one-loop one-, two- and three-point Feynman integrals with
arbitrary tensor rank and arbitrary powers of propagators. Working in
orthogonal and parallel space of momentum configuration \cite{kreimer}, this
procedure allows one to reduce the one-loop tensor integrals to a basis of
simpler integrals without using the Passarino-Veltman procedure. This
procedure has been implemented in the new version of the \xloops\ program,
called \xloops-\ginac.

Unlike the original \xloops, \xloops-\ginac\ currently only handles the
one-loop case (and as such competes with existing programs for one-loop
calculations such as those presented in~\cite{hahn}) but it has been designed
with being a prerequisite for two-loop problems in mind, and the authors'
goal is to make \xloops-\ginac\ a powerful tool for one- and two-loop
analytical Feynman integrations.

In sections 2 and 3 we introduce briefly the motivation to rewrite \xloops\ in
\CC, and the \ginac\ library. In section 4 the procedure used to reduce
one-loop two- and three-point integrals is presented. In two appendices 
simple examples of \CC\ programs using the \ginac\ library and the \xloops-\ginac\
interface for one-loop integrals are presented.

\section{\xloops-\ginac: The Motivation}

In the past, the \xloops\ package \cite{xloops} has been developed in a
heterogenous environment: The core routines for transforming Feynman graphs
into the basic integrals and for analytic integration are implemented in the
language of the \maple~\cite{maple} computer algebra system, the graphical
user interface is written in Tcl/Tk~\cite{tcltk}, and the numerical
integration is done by C programs that are generated and compiled at
run-time by \xloops.

This way of implementation has a couple of drawbacks:

\begin{itemize}
\item While \maple\ and other computer algebra systems provide sophisticated mathematical capabilities,
 it is not suited as an environment for developing large applications such
 as \xloops, as it was not primarily intended as a programming language
 and only offers limited support for modern software engineering. For
 example, the only structured data type in \maple\ is the list, the
 distinction between local and global variables is not consistently
 enforced by all language constructs, and tools such as
 debuggers are very rudimentary.
\item Different versions of \maple\ are in places incompatible with respect
 to the language. This leads to \xloops\ having to provide two versions of
 all program parts written in \maple\, one for \maple\ V Release 1 and one
 for Release 3. With Release 4 and later releases \xloops\ cannot currently
 be run. Not only does the parallel development of multiple program versions
 require higher maintenance effort, but it is also impractical for the user
 to require the installation of a specific version of \maple\ just to run
 \xloops.
\item \maple\ is a commercial system, which prevents \xloops\ from getting
 a wide reach of distribution, especially in academic institutions.
\item The communication between different program parts is difficult because
 \maple\ has insufficient support for embedding programs written in other
 languages, it requires higher efforts for the conversion
 of data and, in some cases, for the multiple storing of redundant
 information (for example, the information about graph topologies has
 to be duplicated for the calculation and for the graphical interface).
\item In general, it is more difficult to maintain and develop a heterogenous
 program package. For the programmer it is necessary to become acquainted
 with three different environments (\maple, Tcl/Tk, and C).
\end{itemize}

The listed drawbacks and the observation of bugs related to the internal
structure of \xloops\ led to the conclusion that the program had reached a
state in which further development was almost impossible (approx.\ 15000
lines of badly documented source code, in some places very obscure
programming techniques like identifiers whose meaning depends on
capitalization). Thus, the decision was made to rewrite \xloops\ from
scratch, putting it on more solid grounds.

In particular, it was decided that the new version of \xloops\ should be
written in one uniform programming language. As the essential part of
\xloops\ are the analytical calculations, traditional computer algebra
systems (\maple, \mathematica~\cite{mathematica}, \reduce~\cite{reduce} and
\mupad~\cite{mupad}) were envisaged at first. But all these share more
or less the same deficiencies as \maple, especially the low suitability of
the built-in language for the development of large systems. It was
therefore decided to use an established programming language as the
foundation and extend it by the required algebraic capabilities. During
the development of \xloops\ it was noted that only a small subset of the
mathematical functions provided by \maple\ are actually needed. These are:
\begin{itemize}
\item complex arithmetic with arbitrary precision;
\item simple manipulation of symbolic expressions, like expansion, collection,
 substitution of variables;
\item simplification of rational functions;
\item symbolic differentiation;
\item special functions like polylogarithms and the Gamma function;
\item Laurent series expansion of expressions containing these functions;
\item solving systems of linear equations for the analysis of the tensor
 structure of the integrals;
\item handling expressions containing elements of some particular
 non-com\-mu\-ta\-tive algebras such as Clifford and SU(3) Lie algebras;
\item numeric integration.
\end{itemize}
Features that are particularly \emph{not} required are:
\begin{itemize}
\item symbolic integration since only few master integrals need to be really integrated and they will be  done by hand;
\item calculation of limits;
\item treatment of domains and assumptions about the range of values of
 variables.
\end{itemize}
Due to the following reasons, \CC\ has been chosen as the programming language:
\begin{itemize}
\item \CC\ is officially standardized~\cite{cstd}, so fewer complications
 are expected from the future development of the language.
\item \CC\ allows to write down symbolic expressions in their natural
 mathematical notation by means of operator overloading (e.g. \verb|4*a+b|
 instead of something like \verb|add(mul(4,a),b)|).
\item \CC\ is available for virtually all computer platforms. In particular,
 there are free compilers for the Unix systems predominant in the  academic
 area.
\item There is a large assortment of development tools available like
 powerful source-level debuggers and systems for version control and
 documentation.
\item There is also a large number of existing libraries, especially for
 arbitrary precision arithmetics.
\item As a compiled language, \CC\ is also suitable for numeric integration.
\end{itemize}

Based on \CC\, we developed a system called "\ginac"\footnote{an acronym
for "\ginac\ is not a Computer Algebra System"} that is primarily aimed at
the re-im\-ple\-men\-ta\-tion of \xloops, but is also suited for the development of
other systems that integrate algebraic and numeric calculations with user
interfaces for methods of data acquisition.

\section{\ginac: A New Programming Environment For \xloops}

\ginac~\cite{ginac} is a \CC\ library for handling symbolic mathematical
expressions \footnote{\url{http://www.ginac.de}}. Some of the features of the library are

\begin{itemize}
\item complex arithmetic with arbitrary precision, based on the \cln\ library
 by Bruno Haible~\cite{cln};
\item manipulation of symbolic expressions;
\item normalization of rational functions;
\item matrices and systems of linear equations;
\item numerous special functions (trigonometric and hyperbolic functions,
 exponential functions, logarithms, Gamma und polygamma functions);
\item symbolic differentiation;
\item series expansion of functions (Taylor and Laurent series);
\item Clifford and SU(3) color algebras (this is a recently added feature in
 GiNaC, making it especially suited for applications in particle physics);
\item it is Open Source, licensed under the GNU General Public License. This
 means that it is not only freely available but also free in the sense that
 users have access to the sources and are allowed to modify or extend the
 library and to redistribute the modified version. This is in sharp contrast
 to most other CAS that place heavy restrictions on their legal use.
\end{itemize}

\ginac\ is designed in an object-oriented fashion. The central class of
\ginac\ is the class \verb|ex| that stores a symbolic expression. Strictly
speaking, \verb|ex| only represents a "smart" pointer to the real
expression which is stored as a tree whose nodes are objects subclassed
from the abstract base class \verb|basic|. The operators \verb|+|,
\verb|-|, \verb|*| und \verb|/| are overloaded to simplify the creation of
expressions in the program code.

\begin{table}
\begin{center}
\begin{tabular}{rp{9cm}}
\toprule
\verb|numeric| &
 real and complex numbers (integers like $ 42 $, exact fractions like
 $ \frac{2}{3} $ and floating point numbers like $ 7.319 $), based on \cln\
\\
\verb|symbol| &
 algebraic symbols
\\
\verb|constant| &
 constants like $ \pi $ which are treated similarly to symbols but also have
 a predefined numerical value
\\
\verb|add| &
 sums of the form $ \displaystyle{\sum_{i=1}^{n}c_i x_i}, $ for
 $ c_i \in \mathbb{C} $ and arbitrary non-numeric expressions $ x_i $
\\
\verb|mul| &
 product of the form $ \displaystyle{\prod_{i=1}^{n}{x_i}^{c_i}}, $ for
 $ c_i \in \mathbb{C} $ and arbitrary non-numeric expressions $ x_i $
\\
\verb|power| &
 arbitrary expressions of the form $ x^y $
\\
\verb|pseries| &
 compact representation of Laurent and Taylor series (only contains the
 series coefficients, the expansion variable, and the expansion point)
\\
\verb|function| &
 mathematical functions like $\sin()$ und $\cos()$, where the individual
 functions are not implemented as subclasses of \texttt{function} but are
 distinguished by a function index
\\
\verb|lst| &
 lists of expressions
\\
\verb|matrix| &
 matrices
\\
\verb|relational| &
 equations and unequations
\\
\bottomrule
\end{tabular}
\end{center}
\caption{The most important \ginac\ classes}
\label{classes}
\end{table}

Table~\ref{classes} gives an overview of the most important subclasses of
\verb|basic|. The classes can be categorized into the "atomic" classes
\verb|numeric|, \verb|symbol| and \verb|constant| which are at the leaves
of an expression tree, and the container classes (all others) which themselves
contain expressions. The representation of sums and products with numeric
coefficients and powers was chosen for reasons of efficiency (see~\cite{geddes}
and~\cite{alex}).

A detailed description of the internal functionality of the basic classes
and methods of \ginac\ is given in~\cite{ginac} and~\cite{alex}.

\section{One-Loop One-, Two-, And Three-Point Integrals}

In this section we present our algorithm for tensor reduction, which was
introduced by Collins~\cite{collins} and further developed by Kreimer~\cite{kreimer,kreimer2}.
A similar approach was used in a subsequent paper by Ghinculov and Yao~\cite{ghi}.
The advantages of this approach compared to other procedures like the
Passarino-Veltman procedure, especially with regard to future two-loop
applications, are outlined in~\cite{alex}.

To regularize UV-divergences, we are working in $D=4-2\epsilon$ dimensional
space-time. The main idea is to split the space of integration in a parallel
and an orthogonal space. We define the parallel space to be the linear span
of the $n$-external momenta $q_{i_\mu}$ ($i=1,...,n$) involved in the integrand.
This parallel space has a finite dimension $J \leq D$. The remaining $D-J$
dimensions span an orthogonal complement of the parallel space called the
orthogonal space \cite{kreimer}.

Once an explicit configuration of external momenta is chosen, the dimension of the parallel space $J$ is known, and the scalar products are written explicitly in terms of the components of the external momenta
\begin{equation}
\begin{split}
l^2 & = l_0^2-l_1^2-\dots-l_{J-1}^2-l_\perp^2, \\
l\cdot q_i & = l_0 q_{i_0}-l_1 q_{i_1}-\dots-l_{J-1} q_{i_{J-1}}.
\end{split}
\end{equation}
A general one-loop integral can be written as
\begin{equation}
\int d^D l \, F(l^2,\{l\cdot q_i\}) = \frac{2\pi^{\frac{D-J}{2}}}{\Gamma((D-J)/2)}\int^\infty_{-\infty} \! dl_0\dots\int^\infty_{-\infty} \! d l_{J-1}\int_0^\infty \! dl_\perp\,l_\perp^{D-J-1}\,F(l_\perp, l_0, \dots, l_{J-1}).
\label{eq0}
\end{equation} 
In the usual Passarino-Veltman approach, to keep an explicitly covariant form, one expands tensor integrals of the type
\begin{equation}
T^{(n)}_{\mu_1 \dots \mu_N} = \int\!\! d^D l \, \frac{l_{\mu_1} \dots l_{\mu_N}}{\prod_{i=1}^{n}\left(\left(l+q_i\right)^2-m_i^2+i \rho\right)}
\end{equation} 
in a basis of Lorentz tensors constructed from the metric tensor $g_{\mu\nu}$ and the external momenta $q_{i_\mu}$.
Since $T^{(n)}_{\mu_1
\dots \mu_N}$ is symmetric, one can re-group indices
belonging to the parallel space and to the orthogonal space together
\begin{equation}
T^{(n)}_{\mu_1 \dots \mu_N} = T^{(n)}_{\mu_J \dots \mu_{D}\underbrace{\underbrace{\hbox{\scriptsize 0\dots 0}}_{p_0}\underbrace{\hbox{\scriptsize 1\dots 1}}_{p_1} \dots \underbrace{\hbox{\scriptsize J-1\dots J-1}}_{p_{J-1}}}_{p_{\|}}}
\end{equation}
where $p_0, p_1, \dots$ denote the numbers of indices for the 0-, 1-, \dots
components. The indices $\mu_J$, \dots $\mu_{D}$ in the orthogonal space
have to result in a symmetric combination of metric tensors in the
orthogonal space $g_{\mu\nu}^\perp$
\begin{equation}
T^{(n)}_{\mu_1 \dots \mu_N} = \frac{(-1)^{\frac{D-J}{2}}}{K}\left(g^\perp_{\mu_J\mu_{J+1}} \dots g^\perp_{\mu_{D-1}\mu_D}\right)_{\it{symm.}}T^{(n)}_{(p_0 \dots p_{J-1}p_\perp)}.
\end{equation}
with $N=p_0+...+p_{J-1}+p_\perp$.
The normalization can be derived by looking at contractions with
$g_{\mu\nu}^\perp$ and observing that
\begin{equation}
g^\perp_{\mu_J\mu_{J+1}}g^{\perp \mu_J\mu_{J+1}} = D-J .
\end{equation}
One obtains
\begin{equation}
K = \prod_{i=J}^{\frac{D-J-2}{2}}(D-J+2i).
\end{equation}
The coefficients needed for the calculation of a specific component of a general one-loop tensor integral therefore have the form
\begin{equation}
T^{(n)}_{(p_0 \dots p_{J-1}p_\perp)} = \int\!\! d^D l \, \frac{l_0^{p_0} \dots l_{J-1}^{p_{J-1}}l_\perp^{p_\perp}}{\prod_{i=1}^{n}\left(\left(l+q_i\right)^2-m_i^2+i \rho\right)}.
\label{eq01}
\end{equation}
In the next sections, we present an algorithm to calculate $T^{(n)}_{(p_0 \dots p_{J-1}p_\perp)}$ for the one-loop two- and three-point tensor integrals. In the rest of this paper, we call $T^{(n)}_{(p_0 \dots p_{J-1}p_\perp)}$ tensor integrals.
\subsection{One-Loop Two-Point Tensor Integrals}

In this section, the algorithm for an automatic calculation of one-loop
two-point tensor integrals is presented.
\begin{figure}[ht]
\begin{center}
\mbox{\epsfysize = 3cm \epsffile{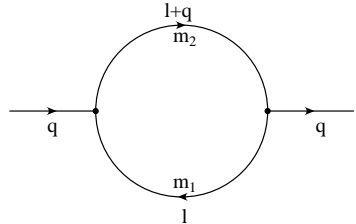}}
\end{center}
\vspace{-8mm}
\caption{Notation for two-point functions}
\label{fig1}
\end{figure}
We first consider the case where $q^2$ is timelike, $q^2>0$. Then one can
choose a reference frame where $q_\mu$ = $(q_0,0,0,0)$. The general integral
for a tensor Feynman diagram shown in Fig.\ref{fig1} has the form
\begin{equation}
I_{t_1t_2}^{(2)\,ij}(q^2) = \intddl \, \frac{l_0^i\,l_\perp^j}{P_1^{t_1} P_2^{t_2}}
\label{eq1}
\end{equation} 
with
\begin{equation}\label{eq2}
\begin{split}
P_1 & = (l_0+q_0)^2 - l_\perp^2 - m_1^2 + i\rho, \\
P_2 & = l_0^2 - l_\perp^2 - m_2^2 + i\rho,
\end{split}
\end{equation}
where $l_0$ and $l_\perp$ span the parallel and orthogonal subspaces
respectively and the integral vanishes unless $j$ is even. In the spacelike
and lightlike cases where $q^2<0$ or $q^2=0$ one can choose a reference
frame where $q_\mu$ = $(0,q_{01},0,0)$ or $q_\mu$ = $(q_{01},q_{01},0,0)$
respectively and the integral space can be split into a two-dimensional
parallel and a $D-2$ dimensional orthogonal subspace. We will consider these
cases later in section~\ref{spacelike}. A genuine one-loop integral has $t_1=t_2=1$, but the more general case is needed in the case of the reduction of integrals with more than one loop.

The strategy now is to express $I_{t_1t_2}^{(2)\,ij}(q^2)$ as a polynomial of
simpler integrals. It turns out that the usual scalar one- and two-point
integrals $A_0$ and $B_0$ (in Passarino-Veltman notation \cite{pv,pvh}) are sufficient. This expansion is always possible except for some
special cases that we will consider later.

Firstly consider the general case where $q_0 \neq 0$. We express the
numerator of the integral in Eq.(\ref{eq1}) as a function of $P_1$ and
$P_2$. From Eq.(\ref{eq2}) we get:
\begin{equation}\label{eq3}
\begin{split}
l_0 & \rightarrow l_0(P_1,P_2,q_0,m_1,m_2) = {\frac{1}{2q_0}}(P_1 - P_2 - C_1), \\
l_\perp^2 & \rightarrow l_\perp^2(P_1,P_2,q_0,m_1,m_2) = l_0^2 - P_2 + C_2, \\
C_1 & = q_0^2 - m_1^2 + m_2^2, \\
C_2 & = -m_2^2 + i\rho.
\end{split}
\end{equation}
Inserting Eq.(\ref{eq3}) into Eq.(\ref{eq1}) and expanding the numerator of
the integrand, one obtains
\begin{equation}
I_{t_1t_2}^{(2)\,ij}(q^2) = \sum_{n,m}^{i+j} C_{nm} \intddl \, P_1^{n-t_1} \, P_2^{m-t_2}
\end{equation}
with $C_{nm}$ being simple functions of $q_0$, $m_1$, $m_2$, and 
\begin{equation}
\intddl \, P_1^{n-t_1}\,P_2^{m-t_2}=\left\{ \begin{array}{ll}
0 & \mbox{if $n-t_1\geq 0$ and $m-t_2 \geq 0$} ,\\ \\
\displaystyle \intddl \, \frac{1}{P_1^{t_1-n}\,P_2^{t_2-m}}& \mbox{if $n-t_1 < 0$ and $m-t_2 < 0$} ,\\ \\
\displaystyle \intddl \, \frac{P_1^{n-t_1}}{P_2^{t_2-m}}& \mbox{if $n-t_1 \ge 0$ and $m-t_2 < 0$} ,\\ \\
\displaystyle \intddl \, \frac{P_2^{m-t_2}}{P_1^{t_1-n}}& \mbox{if $n-t_1 < 0$ and $m-t_2 \ge 0$} .
\end{array}\right.
\label{eq5}
\end{equation}
We see that the second case actually corresponds to a scalar two-point
function. For the last two cases, from Eq.(\ref{eq2}) one can insert
\begin{equation}
P_1 = P_1(P_2,l_0,q_0) = 2 l_0 q_0 + P_2 + C_1,
\end{equation}
or
\begin{equation}
P_2 = P_2(P_1,l_0,q_0) = P_1 - 2 l_0 q_0 - C_1
\end{equation}
and expand the numerator of the integrands in Eq.(\ref{eq5}). Then
Eq.(\ref{eq2}) can be completely reduced to the one-point functions
\begin{equation}
I^{(1)\,i}_t = \intddl \, \frac{(l^2)^{\frac{i}{2}}}{(l^2-m^2+i \rho)^t}.
\end{equation}

\subsubsection{The case $q_0 = 0$}
If $q_0 = 0$, Eq.(\ref{eq5}) must be rewritten using
\begin{equation}\label{eq6}
\begin{split}
P_1 & = l_0^2 - l_\perp^2 - m_1^2 + i\rho, \\
P_2 & = l_0^2 - l_\perp^2 - m_2^2 + i\rho.
\end{split}
\end{equation}
If $m_1 = m_2$ then $P_1 = P_2$ and $I_{t_1t_2}^{(2)\,ij}(q^2)$ has the
simple form
\begin{equation}
I_{t_1t_2}^{(2)\,ij}(q^2) = \intddl \, \frac{l_0^i\,l_\perp^j}{P_1^{t_1+t_2}}
\end{equation}
that is actually the one-loop one-point function.

If $m_1 \neq m_2$, one performs partial fraction decomposition and finds
\begin{equation}
\begin{split}
I_{t_1t_2}^{(2)\,ij}(q^2) & = \intddl \, \frac{l_0^i\,l_\perp^j}{P_1(m_1^2)^{t_1}\,P_2(m_2^2)^{t_2}} \\
 & = \frac{1}{(t_1-1)!}\frac{d^{t_1-1}}{d(m_1^2)^{t_1-1}}\left(\frac{1}{(m_1^2-m_2^2)^{t_2}} \intddl \, \frac{l_0^i\,l_\perp^j}{P_1(m_1^2)} \right) + \\
 & \quad\,\, \frac{1}{(t_2-1)!}\frac{d^{t_2-1}}{d(m_2^2)^{t_2-1}}\left(\frac{1}{(m_2^2-m_1^2)^{t_1}} \intddl \, \frac{l_0^i\,l_\perp^j}{P_2(m_2^2)} \right),
\end{split}
\end{equation}
which is a combination of one-loop one-point functions.

\subsection{One-Loop Three-Point Tensor Functions}
\begin{figure}[ht]
\begin{center}
\mbox{\epsfysize = 3cm \epsffile{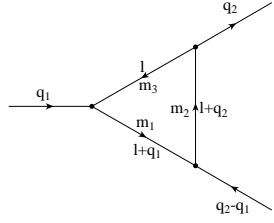}}
\end{center}
\vspace{-8mm}
\caption{Notation for three-point functions}
\label{fig2}
\end{figure}
The notation we use for one-loop three-point functions is shown in
Fig.\ref{fig2}. We are working in the frame of reference where the external
momentum configuration is
$q_{1_\mu}=(q_{10},0,0,0)$, $q_{2_\mu}=(q_{20},q_{21},0,0)$. The parallel
space is now two-dimensional and the general form of the one-loop three-point
tensor function is
\begin{equation}
I_{t_1\,t_2\,t_3}^{(3)\,ijk} = \intddl \, \frac{l_0^i\,l_1^j\,l_\perp^k}{P_1^{\, t_1}\,P_2^{\, t_2}\,P_3^{\, t_3}}
\label{eq9}
\end{equation}
with
\begin{equation}\label{eq10}
\begin{split}
P_1 & = l_0^2-l_1^2-l_\perp^2+2\,l_0\,q_{10}+q_{10}^2-m_1^2+i\rho, \\
P_2 & = l_0^2-l_1^2-l_\perp^2+2\,l_0\,q_{20}-2\,l_1\,q_{21} + q_{20}^2 - q_{21}^2 - m_2^2 + i\rho, \\
P_3 & = l_0^2-l_1^2-l_\perp^2-m_3^2+i\rho
\end{split}
\end{equation}
where \{$l_0$, $l_1$\} and $l_\perp$ span the parallel and orthogonal
subspaces, respectively.

\subsubsection{The general case}
Firstly we consider the case where $q_{10}\neq 0$ and $q_{21}\neq 0$. Then we are
always able to express $l_0$, $l_1$, $l_\perp$ in terms of $P_1$, $P_2$, $P_3$:
\begin{equation}\label{eq11}
\begin{split}
l_0 & = {\frac{1}{2 q_{10}}}(P_1-P_3-c_{00}), \\
l_1 & = {\frac{1}{2 q_{10} q_{21}}}\left[q_{20}\left(P_1-P_3-\left(q_{10}^2-m_1^2+m_3^2\right)\right) +q_{10}\left(P_3-P_2\right)+q_{10}\left(m_3^2-m_2^2+q_{20}^2-q_{21}^2\right)\right] \\
    & = {\frac{1}{2 q_{21}}}\left[c_{10}l_0+(P_3-P_2)+c_{11}\right], \\
l_\perp^2 & = l_0^2-l_1^2-P_3-c_{20}
\end{split}
\end{equation}
with 
\begin{equation}
\begin{split}
c_{00} & = q_{10}^2-m_1^2+m_3^2, \\
c_{10} & = 2 q_{20}, \\
c_{11} & = m_3^2-m_2^2+q_{20}^2-q_{21}^2, \\
c_{20} & = m_3^2-i\rho.
\end{split}
\end{equation}
Again, as in the case of two-points functions, one substitutes
Eq.(\ref{eq11}) into Eq.(\ref{eq9}) and obtains a combination of scalar
three-point functions
\begin{equation}
I_{t_1 t_2 t_3}^{(3)\,ijk} = \sum_{m,n,r}^{i+j+k} C_{mnr} \intddl \, P_1^{(m-t_1)} P_2^{(n-t_2)} P_3^{(r-t_3)}
\label{eq12}
\end{equation}
with $C_{mnr}$ being simple functions of masses and components of external
momenta. More explicitly, the integrand on the right hand side of
Eq.(\ref{eq12}) contains terms like
\begin{equation}
\frac{1}{P_1^{n_1}\,P_2^{n_2}\,P_3^{n_3}}, \quad \frac{P_i^{n_i}\,P_j^{n_j}}{P_k^{n_k}}, \quad \text{and} \quad \frac{P_i^{n_i}}{\,P_j^{n_j}\,P_k^{n_k}}
\end{equation}
with $i\neq j\neq k$ and positive $n_i$. Other possible combinations lead
to a vanishing integral.

The first group of terms can be obtained from derivatives of scalar
three-point functions. For the last two cases, using Eq.(\ref{eq10}) one can
expand the numerator in terms of propagators $P_{n_j}$ in the denominator.
This expansion is always possible and reduces the second group of terms to
one-point functions. Similarly the third group of terms can be reduced to
one-loop two-point functions. Note that in this step we meet two kinds of
one-loop two-point functions. The first kind is the one-loop two-point
function with one parallel dimension that was already found in the previous
section. The second kind is the one-loop two-point function with two
parallel dimensions of the internal momentum $l$ that is not trivial and
will be given in the section 4.2.3 as a separate case.

\subsubsection{The case $q_{10}\,q_{21} = 0$}
In this case the expansions in Eq.(\ref{eq11}) cannot be used. However,
Eq.(\ref{eq9}) can be reduced by partial fraction decomposition. First
consider the case $q_{10} = 0$ with arbitrary values of $q_{21}$.

\subsubsection*{4.2.2.1 The case $q_{10} = 0$}
Eq.(\ref{eq10}) simplifies to
\begin{equation}\label{eq13}
\begin{split}
P_1 & = l_0^2-l_1^2-l_\perp^2-m_1^2+i\rho, \\
P_2 & = l_0^2-l_1^2-l_\perp^2+2 l_0 q_{20}-2 l_1 q_{21} + q_{20}^2 - q_{21}^2 - m_2^2 + i\rho, \\
P_3 & = l_0^2-l_1^2-l_\perp^2-m_3^2+i\rho.
\end{split}
\end{equation}

If $m_1 \neq m_3$, partial fraction decomposition leads to a separation into
terms which contain only two propagators
\begin{equation}
I_{t_1 t_2 t_3}^{(3)\,ijk} = \prod_{f=1}^3 \left( \frac{1}{(t_f-1)!}\frac{d^{t_f-1}}{d(m_f^2)^{t_f-1}} \right)
  \left\{ \frac{1}{m_1^2-m_3^2} \left( \intddl \, \frac{l_0^i\,l_1^j\,l_\perp^k}{P_1(m_1^2)\,P_2(m_2^2)} - \intddl \, \frac{l_0^i\,l_1^j\,l_\perp^k}{P_2(m_2^2)\,P_3(m_3^2)} \right) \right\}.
\label{eq14}
\end{equation}
If $m_1 = m_3$, two propagators are equal and it is sufficient to calculate
\begin{equation}
I_{t_1\,t_2\,t_3}^{(3)\,ijk} = \intddl \, \frac{l_0^i\,l_1^j\,l_\perp^k}{P_1^{t_1+t_3}(m_1^2)\,P_2^{t_2}(m_2^2)}.
\label{eq15}
\end{equation}
The three-point integrals are then reduced to combinations of two-point
integrals with two parallel dimensions. We will treat this class of two-point
functions in section 4.2.3.

\subsubsection*{4.2.2.2 The case $q_{10} \neq 0$ and $q_{21} = 0$}
In this special case, the integrals collapse from two parallel dimensions to
one and Eq.(\ref{eq10}) reads
\begin{equation}\label{eq16}
\begin{split}
P_1 & = (l_0+q_{10})^2-l_\perp^{\prime 2}-m_1^2+i\rho, \\
P_2 & = (l_0+q_{20})^2-l_\perp^{\prime 2}- m_2^2 + i\rho, \\
P_3 & = l_0^2-l_\perp^{\prime 2}-m_3^2+i\rho
\end{split}
\end{equation}
where the components $l_1$ and $l_\perp$ can be combined into $l_\perp^{\prime}$ to form a new $D-1$ dimensional orthogonal subspace with
\begin{equation}
l_\perp^{\prime 2} = l_1^2 + l_\perp^2.
\end{equation}
Then the integral can be reduced to a simpler integral with only one parallel dimension 
\begin{equation}
I_{t_1\,t_2\,t_3}^{(3\,)i,j+k} \,=\, \intddl \, \frac{l_0^i \,l_\perp^{j+k}}{P_1^{t_1} P_2^{t_2} P_3^{t_3}}.
\label{eq17}
\end{equation}
Again, one can use the same procedure as in the general case 
\begin{equation}
\begin{split}
l_0 \rightarrow l_0(P_1,P_3,m_i,q_{10}) & = \frac{1}{2 q_{10}}(P_1-P_3+m_1^2-m_3^2-q_{10}^2), \\
l_\perp \rightarrow l_\perp(P_1,P_3,m_i,q_{10}) & = l_0^2-P_3-m_3^2 + i \rho
\end{split}
\end{equation}
that reduces Eq.(\ref{eq17}) to a form similar to Eq.(\ref{eq12}).

\subsubsection{The two-point integral with two parallel dimensions  $J^{ijk}_2$}\label{spacelike}
In the preceding section we have shown that general one-loop three-point
tensor functions can be reduced to the usual scalar integrals and one new
two-point function corresponding to a tensor component in a two-dimensional
parallel space. Explicitly, this integral reads
\begin{equation}\label{eq19}
\begin{split}
J^{ijk}_2 & = \intddl \, \frac{l_0^i\;l_1^j\;l_\perp^k}{[(l_0+q_{10})^2-l_1^2-l_\perp^2-m_1^2+i \rho]^{t_1}\;[(l_0+q_{20})^2-(l_1+q_{21})^2-l_\perp^2-m_2^2+i \rho]^{t_2}} \\
 & = \intddl \, \frac{(l_0-q_{10})^i\;l_1^j\;l_\perp^k}{[l_0^2-l_1^2-l_\perp^2-m_1^2+i \rho]^{t_1}\;[(l_0+Q_0)^2-(l_1+Q_1)^2-l_\perp^2-m_2^2+i \rho]^{t_2}}
\end{split}
\end{equation}
with $Q_0=q_{20}-q_{10}$, $Q_1=q_{21}$; $Q_\mu=q_{2_\mu}-q_{1_\mu}$. If $Q_1=0$, this integral reduces to the one-loop
two-point function in one-dimensional parallel space as found in the
previous section. If, on the other hand, $Q_1 \neq 0$, one can always  find a Lorentz boost which
transforms into a reference frame where the transformed 4-momentum $Q_\mu^\prime$ has either only one non-zero component ($Q_0^\prime$ or $Q_1^\prime$ if $Q$ is timelike or spacelike) or where $Q_0^\prime= Q_1^\prime$ if $Q$ is lightlike. The loop momentum has to be boosted accordingly which, however, modifies only the numerator of the integrand in Eq.(\ref{eq19}). Explicitly, consider the boost
\begin{equation}
\begin{pmatrix} l_0 \\ l_1 \end{pmatrix} = \begin{pmatrix} \gamma & \gamma\beta \\ \gamma\beta & \gamma \end{pmatrix} \begin{pmatrix} l_0^\prime \\ l_1^\prime \end{pmatrix}.
\label{eq20}
\end{equation}
Then the three sub-cases are treated as follows.

\subsubsection*{4.2.3.1 The timelike case $Q_0^2-Q_1^2 > 0$}
In this case, under the transformation in Eq.(\ref{eq20}) the integral will
be reduced to one-loop two-point functions with a one-dimensional parallel space
as found in the previous section:
\begin{equation}
J^{ijk}_2 = \int\!\! d^D l^{\prime} \, \frac{\left[\gamma\left(l_0^{\prime}+\beta \,l_1^{\prime}\right)-q_1\right]^i\;\left[\gamma\left(\beta\, l_0^{\prime}+l_1^{\prime}\right)\right]^j\;l_\perp^{\prime \,k}}{\left[l_0^{\prime\,2}-l_1^{\prime\,2}-l_\perp^{\prime\,2}-m_1^2+i \rho\right]^{t_1}\,\left[(l_0^{\prime}+P)^2-l_1^{\prime\,2}-l_\perp^{\prime\,2}-m_2^2+i \rho\right]^{t_2}}
\end{equation}
with $P = \sqrt{Q_0^2-Q_1^2}$, ${\textstyle \gamma = Q_0/P}$ and ${\textstyle \beta = Q_1/Q_0}$.

\subsubsection*{4.2.3.2 The spacelike case, $Q_0^2-Q_1^2 < 0$}
In this case the boost with $P = \sqrt{Q_1^2-Q_0^2}$, ${\textstyle \gamma = Q_1/P}$ and
${\textstyle \beta = Q_0/Q_1}$ transforms to a reference frame in which the integral reads
\begin{equation}
J^{ijk}_2 = \int\!\! d^D l^{\prime} \, \frac{\left[\gamma\left(l_0^{\prime}+\beta\, l_1^{\prime}\right)-q_1\right]^i\;\left[\gamma\left(\beta\, l_0^{\prime}+l_1^{\prime}\right)\right]^j\;l_\perp^{\prime \,k}}{\left[l_0^{\prime\,2}-l_1^{\prime\,2}-l_\perp^{\prime\,2}-m_1^2+i \rho\right]^{t_1}\,\left[l_0^{\prime\,2}-(l_1^{\prime}+P)^2-l_\perp^{\prime\,2}-m_2^2+i \rho\right]^{t_2}}
\end{equation}
The components $l_0^{\prime}$ and
$l_\perp^\prime$ can be combined to form a new $D-1$ dimensional orthogonal
subspace while $l_1$ spans the parallel subspace. Using the same procedure as
in the previous sections one can  reduce the integral completely to the
scalar one- and two-point functions.

\subsubsection*{4.2.3.3 The lightlike case, $Q_0^2 - Q_1^2 = 0$}
In this case the transformation in Eq.(\ref{eq20}) is singular and the
integral $J^{ijk}_2$ becomes
\begin{equation}
J^{ijk}_2 = \intddl \, \frac{(l_0-q_1)^i\;l_1^j\;l_\perp^k}{[l_0^2-l_1^2-l_\perp^2-m_1^2+i \rho]^{t_1}\,[(l_0+Q_0)^2-(l_1+Q_0)^2-l_\perp^2-m_2^2+i \rho]^{t_2}}.
\label{eq21}
\end{equation}
By solving the system of equations
\begin{equation}
\begin{split}
P_1 & = l_0^2-l_1^2-l_\perp^2-m_1^2+i\rho ,\\
P_2 & = (l_0+Q_0)^2-(l_1+Q_0)^2-l_\perp^2-m_2^2+i \rho ,
\end{split}
\end{equation}
one obtains
\begin{equation}\label{eq22}
\begin{split}
l_\perp^2 & = l_0^2-l_1^2-P_1-m_1^2+i\rho, \\
l_1 & = \frac{P_1-P_2+m_2^2-m_1^2}{2 Q_0}+l_0.
\end{split}
\end{equation}
Inserting Eq.(\ref{eq22}) into Eq.(\ref{eq21}) and expanding the numerator
of the integrand, the integral will be reduced to scalar one-point
functions and a simpler tensor integral
\begin{equation}
J^{i}_2 = \intddl \, \frac{(l_0)^i}{[l_0^2-l_1^2-l_\perp^2-m_1^2+i \rho]^{t_1}\,[(l_0+Q_0)^2-(l_1+Q_0)^2-l_\perp^2-m_2^2+i \rho]^{t_2}} .
\end{equation}
The explicit calculation of this integral is given in \cite{lars}.

This completes the description of our algorithm for tensor reduction. We did
not reproduce explicit expressions for the basic scalar integrals in this
paper since these can be found in the literature~\cite{kreimer,pvh}.

\section{Conclusion}
Due to the limitations of Maple and the internal structure of \xloops\ we
decided to rewrite \xloops\ from scratch, based on \ginac, a \CC\ library to
replace Maple as an algebraic programming environment \cite{ginac}. An
efficient algorithm for one-loop one-, two- and three-point tensor reduction
was also successfully implemented. At this stage of the project, a package
for calculating one-loop one-, two- and three-point tensor integrals is
available and can be downloaded from \url{http://wwwthep.physik.uni-mainz.de/~xloops}.
Like GiNaC, it is distributed under the terms of the GNU General Public License.

As the next step, we plan to rewrite the module for two-loop one-, two- and
three-point integrals and to completely recode \xloops\ in \CC\ using the
\ginac\ library, providing a package for doing particle physics in a
homogenous \CC\ environment.

\section*{Acknowledgements}
Part of this work is supported by the DFG-Forschungsproject "KO 1069/6-1"
and the "Graduiertenkolleg Eichtheorien --- Experimentelle Tests und
theoretische Grundlagen" at the University of Mainz. The authors would like
to thank J\"urgen K\"orner, Dirk Kreimer and Hubert Spiesberger for many
fruitful comments and corrections, and Lars Br\"ucher, Alexander Frink and
Richard Kreckel for inspiring discussions.

\section*{Appendix A}
In this section we introduce the definitions for the one-loop integral
functions in \xloops-\ginac. The \xloops-\ginac\ package provides the three
\ginac\ functions \verb|OneLoop1Pt()|, \verb|OneLoop2Pt()|, and \verb|OneLoop3Pt()|,
which can be used in algebraic expressions like any other predefined
\ginac\ function:

\begin{itemize}
\item The one-point function
\begin{equation*}
{\tt OneLoop1Pt(i,m,t,\rho)} = I^{(1)\,i}_t = \intddl \, \frac{(l^2)^{\frac{i}{2}}}{[l^2-m^2+i\rho]^t}.
\end{equation*}
\item The two-point function
\begin{gather*}
{\tt OneLoop2Pt(i,j,q,m_1,m_2,t_1,t_2,\rho)} = I_{t_1t_2}^{(2)\,ij} \\
  = \intddl \, \frac{l_0^i\,l_\perp^j}{[(l_0+q)^2\,-\,l_\perp^2\,-\,m_1^2\,+\,i\rho]^{t_1}\,[l_0^2\,-\,l_\perp^2\,-\,m_2^2\,+\,i\rho]^{t_2}}.
\end{gather*}
\item The three-point function
\begin{gather*}
{\tt OneLoop3Pt(i,j,k,q_{10},q_{20},q_{21},m_1,m_2,m_3,t_1,t_2,t_3,\rho)} = I_{t_1\,t_2\,t_3}^{(3)\,ijk} \\
  = \intddl \, \frac{l_0^i\,l_1^j\,l_\perp^k}{P_1^{\, t_1}\,P_2^{\, t_2}\,P_3^{\, t_3}}
\end{gather*}
with 
\begin{equation*}
\begin{split}
P_1 & = l_0^2-l_1^2-l_\perp^2+2\,l_0\,q_{10}+q_{10}^2-m_1^2+i\rho, \\
P_2 & = l_0^2-l_1^2-l_\perp^2+2\,l_0\,q_{20}-2\,l_1\,q_{21} + q_{20}^2 - q_{21}^2 - m_2^2 + i\rho, \\
P_3 & = l_0^2-l_1^2-l_\perp^2-m_3^2+i\rho.
\end{split}
\end{equation*}
\end{itemize}

As with any \ginac\ function, the arguments and return values of the above
functions are objects of type \verb|ex| so the return values as well as input
parameters can be any symbolic or numeric expression.

In order to illustrate the output of \xloops-\ginac, we give one example
program that calculates and prints out both analytical and numerical results
of the UV-divergent and the finite terms of the integral ${\tt OneLoop2Pt(1,0,q,m_1,m_2,1,1,\rho)}$

\begin{codeblock}
#include <iostream>
#include <ginac/ginac.h>
#include <xloops/xloops.h>
using namespace GiNaC;
using namespace xloops;

int main()
\{
    symbol q("q"), m1("m1"), m2("m2"), eps("eps"), rho("rho");
    ex a = OneLoop2Pt(1, 0, q, m1, m2, 1, 1, rho);
    a = a.series(eps == 0, 4);
    ex a1 = a.coeff(eps, -1).subs(rho == 0);
    ex a2 = a.coeff(eps, 0).subs(rho == 0);
    cout << "Order eps^-1 is " << endl << a1.normal() << endl;
    cout << "Order eps^0 is "  << endl << a2.normal() << endl;
    ex b1 = a.subs(rho==0).subs(m1==80).subs(m2==80).subs(q==100).evalf();
    cout << "Numerical value up to order eps^2 is " << endl << b1 << endl;
    return 0;
\}
\end{codeblock}
The output of this program reads
\begin{codeblock}
Order eps^-1 is
-1/2*I*Pi^2*q

Order eps^0 is
(-1/2*q^(-1)*m2^2-1/2*q+1/2*q^(-1)*m1^2)*(I*Pi^2*q^(-1)*((1/2*q-1/2*q^(-1)*(m1\\
^2-m2^2))*R2ex1(-m2^2,-(1/2*q-1/2*q^(-1)*(m1^2-m2^2))^2)+R2ex1(-m1^2,-(1/2*q+1\\
/2*q^(-1)*(m1^2-m2^2))^2)*(1/2*q+1/2*q^(-1)*(m1^2-m2^2)))+q*(-I*Pi^2*log(Pi)*q\\
^(-1)+Pi^(3/2)*(I*sqrt(Pi)*q^(-1)*(-2*log(2)-Euler)+2*I*log(2)*sqrt(Pi)*q^(-1)\\
+2*I*sqrt(Pi)*q^(-1)))-Pi^3)+1/2*q^(-1)*(-I*Pi^2*m2^2*log(m2^2)-I*Pi^2*log(Pi)\\
*m2^2+Pi^2*(I-I*Euler)*m2^2)-1/2*q^(-1)*(Pi^2*(-I*m1^2*log(m1^2)-I*Euler*m1^2+\\
I*m1^2)-I*Pi^2*log(Pi)*m1^2) 

Numerical value up to order eps^2 is
(-493.48022005446793098*I)*eps^(-1)+(5019.9161138633880865*I)+(-4.6074255521\\
943996428E-15-25944.085010687200793*I)*eps+Order(eps^2)
\end{codeblock}

\section*{Appendix B}
In order to illustrate the use of \xloops-\ginac\ for calculating one-loop
Feynman diagrams, we give one example program that is actually part of the
automated regression tests of the package. It checks that the longitudinal
part of the vacuum polarization in QED vanishes on the one-loop level as
required by gauge invariance, i.e.
\begin{equation}
\label{eqn:ward1}
q_\mu \Pi^{\mu\nu}(q^2) = 0,
\end{equation}
where
\begin{equation*}
\Pi^{\mu\nu}(q^2) = -e^2 \intddl\,\Tr\frac{\gamma^\mu (\lslash+\qslash+m) \gamma^\nu (\lslash+m)}{((l+q)^2-m^2)(l^2-m^2)}.
\end{equation*}
The general tensor structure of $\Pi^{\mu\nu}$ is
\begin{equation*}
\Pi^{\mu\nu} = A g^{\mu\nu} + B \frac{q^\mu q^\nu}{q^2}
\end{equation*}
with functions $A$ and $B$ that, because of~\eqref{eqn:ward1}, satisfy
\begin{equation*}
A+B=0.
\end{equation*}
This expression is obtained by contracting $\Pi^{\mu\nu}$ with $\frac{q_\mu q_\nu}{q^2}$:
\begin{equation}\label{eqn:ward2}
\begin{split}
\Pi^{\mu\nu} \frac{q_\mu q_\nu}{q^2} & = A + B \\
 & = -\frac{e^2}{q^2}\intddl\,\Tr\frac{\qslash (\lslash+\qslash+m) \qslash (\lslash+m)}{((l+q)^2-m^2)(l^2-m^2)}.
\end{split}
\end{equation}
The following program verifies that this expression vanishes for the first
three orders of the regularization parameter:

\label{prg:ginacexample}
\begin{codeblock}
#include <ginac/ginac.h>
#include <xloops/xloops.h>
using namespace GiNaC;
using namespace xloops;

int main(void)
\{
    symbol D("D");
    symbol l("l"), q("q"), m("m"), e("e");
    symbol l0("l0"), lorth("lorth"), eps("eps"), rho("rho");

    scalar_products sp;
    sp.add(l, l, pow(l0, 2) - pow(lorth, 2));
    sp.add(q, q, pow(q, 2));
    sp.add(l, q, l0*q);

    ex I = -pow(e, 2) / pow(q, 2)
         * dirac_slash(q, D)
         * (dirac_slash(l, D) + dirac_slash(q, D) + m * dirac_ONE())
         * dirac_slash(q, D)
         * (dirac_slash(l, D) + m * dirac_ONE());
    I = dirac_trace(I).simplify_indexed(sp);

    ex a;
    for (int i=0; i<3; i++)
        for (int j=0; j<3; j++) \{
            ex c = I.coeff(l0, i).coeff(lorth, j);
            a += c * OneLoop2Pt(i, j, q, m, m, 1, 1, rho);
        \}
    a = a.series(eps == 0, 4);

    ex a1 = a.coeff(eps, -1).subs(rho == 0);
    ex a2 = a.coeff(eps, 0).subs(rho == 0);
    ex a3 = a.coeff(eps, 1).subs(rho == 0);

    cout << "Order eps^-1 is " << a1.expand() << endl;
    cout << "Order eps^0 is " << a2.expand() << endl;
    cout << "Order eps^1 is " << a3.expand() << endl;

    return 0;
\}
\end{codeblock}

\begin{center}
\begin{tabular}{lp{15cm}}
{\bf Line} & {\bf Explanation} \\
\hline \\
1-4   & Include the header files for the \ginac\ and \xloops\ libraries. \\
8-10  & Declaration of all appearing symbols. These are the spacetime dimension
        $D$, the loop and external momenta $l$ and $q$, the mass $m$, the
        parallel and orthogonal space loop components $l_0$ and $l_\perp$,
        the dimensional regularization parameter $\varepsilon$, and the
        infinitesimal imaginary part of the propagator $\rho$. With \ginac\
        it is necessary to specify the name used for printing expressions
        because \CC\ does not provide the names of variables at run-time. \\
12-15 & The possible scalar products of the momenta are expressed in terms of
        $q$, $l_0$ and $l_\perp$ and are registered in a \verb|scalar_products|
        object which is later passed to \verb|simplify_indexed()|:
	$l\cdot l=l_0^2-l_\perp^2, q\cdot q=q^2, l\cdot q=l_0 q.$
	\verb|pow()| is used for exponentiation because the \CC\ operator
        ${\tt\hat{ }}$ has the wrong precedence in relation to the operator
        ${\tt\ast}$. \\
17-21 & The numerator of the integrand of Eq.(\ref{eqn:ward2}) is constructed
        using Clifford algebra objects in a nearly 1-to-1 translation of the
        right-hand side of that equation. Note that GiNaC does not require the
        use of a special operator for non-commutative products here. \\
22    & The trace is taken and the resulting expression which contains
        metric tensors is simplified, inserting the scalar products defined
        above (the result for \verb|I| is $-4 e^2 (l_0^2 + l_\perp^2 + q l_0 + m^2)$).
        Expressions are usually manipulated in the \CC\ oriented notation
        \emph{expression}{\tt $.$}\emph{function}{\tt (}\emph{parameters}{\tt )},
        but the functional notation \emph{function}{\tt (}\emph{expression}{\tt ,}\emph{parameters}{\tt )}
        is also available, as shown. \\
24-29 & The integral is now expressed in terms of the basic \verb|OneLoop2Pt()|
        integral functions by assembling the coefficients of all powers of
        $l_0$ and $l_\perp$. \\
30    & To get the UV divergence as well as the finite part, one needs to take
        the series expansion of the integral at the pole $\varepsilon = 0$. \\
32-34 & The coefficients of the series for the orders $\varepsilon^{-1}$,
        $\varepsilon^0$ and $\varepsilon^1$ are extracted and the limit
        $\rho \rightarrow 0$ is taken by calling the function \verb|subs()|. \\
36-38 & The three coefficients are simplified with \verb|expand()| and printed
        to the standard output stream. When run, this program will output all
        three coefficients as \verb|0|. \\
\\
\hline
\end{tabular}
\end{center}

\end{document}